\documentstyle[amsthm,amssymb,12pt]{article}

\newcommand{\ints}{\Bbb Z}
\newcommand{\reals}{\Bbb R}
\newcommand{\rats}{\Bbb Q}
\newcommand{\nats}{\Bbb N}

\theoremstyle{definition}
\newtheorem{definition}{Definition}

\theoremstyle{remark}
\newtheorem{remark}{Remark} 
\theoremstyle{plain}
\newtheorem{theorem}{Theorem}

\newtheorem{lemma}[theorem]{Lemma}

\newtheorem{corollary}[theorem]{Corollary}
\newtheorem{fact}{Fact}

\begin{document}

\title{Irrational Numbers of Constant Type --- \\
A New Characterization}
\author{ \\
    \\
	Manash Mukherjee \\
	Department of Physics\\
	Virginia Polytechnic Institute and State University\\
	Blacksburg, Virginia  24061  USA\\
	E-mail: manash@phys.vt.edu\\
	Fax: (540) 231-7511\\
	\\
	and \\
	\\
	Gunther Karner\\
	Institut f\"{u}r Kerntechnik und Reaktorsicherheit\\
	Universit\"{a}t Karlsruhe (TH)\\
	Postfach 3640, D-76021 Karlsruhe, Germany\\
	E-mail: karner@irs.fzk.de\\
	Fax: (07247) 82-3718	
}
\date{}	

\parskip .2truecm

\maketitle

\newpage
	\vskip .3truecm
	\centerline{\large \sf Irrational Numbers of Constant Type ---} 
	\centerline{\large \sf A New Characterization}
	\vskip 1truecm
	\centerline{\sf Manash Mukherjee and Gunther Karner}
	\vskip 2.0truecm
\begin{abstract}
	\par \noindent Given an irrational number 
	$\alpha$  
	and a positive integer $m$, 
	the distinct fractional parts of $\alpha, 2\alpha, \cdots, m\alpha$ 
	determine a partition of the interval 
	$[0,1]$.  
	Defining $\displaystyle{d_{\alpha}(m)}$ and 
	$\displaystyle{d'_{\alpha}(m)}$ to be  
	the maximum and minimum lengths, respectively, of the 
	subintervals of the 
	partition corresponding to the integer $m$,
	it is shown that the sequence, 
	$\displaystyle{\left(\frac{d_{\alpha}(m)}
	{d'_{\alpha}(m)} \right)_{m=1}^{\infty}}$, is 
	bounded if and only if $\alpha$ is of constant type.   
	(The proof of this assertion is based on the 
	continued fraction expansion of irrational numbers.)
	In the investigation of certain dynamical systems, this
	result is essential for the formulation of stability
	criteria for orbits of so called quamtum twist maps [3].
\end{abstract}
\newpage
\section{Introduction} \label{sec1}
\par \noindent Let $\alpha$ be a real irrational number, and 
$\alpha - [\alpha] = \{\alpha\}$ be the fractional part of $\alpha$ 
(where $[\cdot]$ is the greatest integer function).  
For $k = 1, 2, \cdots , m$, consider the sequence of distinct 
points $\{k \alpha \}$ in $[0,1]$, arranged in increasing 
order:  
$$0 < \{k_{1}\alpha\} < \cdots < \{k_{j}\alpha\} < 
\{k_{j+1}\alpha\} < \cdots < \{k_{m}\alpha\} < 1$$
\noindent where $1 \leq k_{j} \leq m$ for 
$j = 1, 2, \cdots , m$. 

\par \noindent Let $\displaystyle{d_{\alpha}(m)}$ and 
$\displaystyle{d'_{\alpha}(m)}$ denote, 
respectively, the maximum and minimum lengths of the  
subintervals determined 
by the above partition of $[0,1]$.  Using the continued
fraction expansion of $\alpha$ [see {\sf Section 2}], and 
the {\bf Three Distance Theorem} [see {\sf Section 3}, 
{\bf Theorem \ref{thm1}}], 
we obtain a new characterization of irrational numbers of 
constant 
type (defined as irrationals with bounded partial 
quotients).  We show, in {\bf Theorem \ref{thm2}} [see 
{\sf Section 3}, 
{\bf The Main Theorem}], that the sequence 
$\displaystyle{\left(\frac{d_{\alpha}(m)}
{d'_{\alpha}(m)} \right)_{m=1}^{\infty}}$ is bounded 
if and only if $\alpha$ is an irrational number of constant 
type.

\par \noindent Other characterizations of irrational 
numbers of 
constant type can be found in the survey article by 
J. Shallit [2].  
In the investigation of certain dynamical systems, 
{\bf Theorem \ref{thm2}} is essential for the formulation 
of stability 
criteria for orbits of so called quantum twist maps [3]. 

\noindent 
\section{Basic Properties of Continued 
Fractions} \label{sec2}

\par Throughout this paper, $\nats$, $\ints$, $\rats$, 
$\reals$ denote the natural numbers, 
integers, rational numbers, and real numbers, respectively, 
and $\alpha$ denotes an irrational number.  
Proofs of the following facts can be found in reference 
[4, p. 30]. 

\begin{fact} \label{fa1}
$\alpha \in \reals \setminus \rats$ if and only if $\alpha$ has 
\underline{infinite} (simple) continued fraction expansion:
$$\alpha = [a_{0};a_{1},a_{2},\cdots, a_{n},\cdots] = 
a_{0}+\frac{1}{a_{1}+
\displaystyle{\frac{1}{a_{2}+\ddots}}} $$
\noindent where $a_{0} \in \ints$ and $a_{n} \in \nats$ 
for $n \geq 1$. \hfill $\Box$
\end{fact}

\begin{definition} \label{def1}
An irrational number, $\alpha$, is of constant type 
provided there exists a positive number, $B(\alpha)$,  
such that ${\displaystyle B(\alpha) = \sup_{n \geq 1} 
\left(a_{n}\right) < \infty}$. (See [2])
\hfill $\Box$
\end{definition}

\begin{fact} \label{fa2}
Define integers $p_{n}$ and $q_{n}$ by:
$$\begin{array}{llllll}
 	p_{-1} = 1 & ; & p_{0} = a_{0} & ; & p_{n} = 
 	a_{n}p_{n-1}+p_{n-2} & ,~ n \geq 1  \\
 	q_{-1} = 0 & ; & q_{0} = 1 & ; & q_{n} = 
 	a_{n}q_{n-1}+q_{n-2} & ,~ n \geq 1
 \end{array}$$
Then, for $n \geq 0$, $\gcd(p_{n},q_{n})=1$, and 
$0 < q_{1} < q_{2} < \cdots < q_{n} < q_{n+1} < \cdots$.  
Furthermore, $(q_{n}\alpha - p_{n})$ and 
$(q_{n+1}\alpha - p_{n+1})$ are of opposite sign for all 
$n \geq 0$. \hfill $\Box$
\end{fact}

\noindent {\it Note:}  $\displaystyle{\left( 
\frac{p_{n}}{q_{n}} 
\right)_{n \geq 0}}$ are called principal convergents to 
$\alpha$.

\begin{lemma} \label{lem1}
Let $\eta_{n} \equiv |q_{n}\alpha - p_{n}|$ and 
$\eta_{-1} =1$.  For all $n \geq 0$, 
$\eta_{n-1} = a_{n+1} \eta_{n} + \eta_{n+1}$, 
and hence, $\eta_{n} < \eta_{n-1}$.
\end{lemma}

\begin{proof}
From {\bf Fact \ref{fa2}}, we have 
$$|q_{n-1}\alpha - p_{n-1}| = |(q_{n+1}\alpha - p_{n+1}) - 
a_{n+1}(q_{n}\alpha - p_{n})|$$
The lemma follows from the fact that $a_{n}> 0$ for $n \geq 1$, 
and that $(q_{n}\alpha - p_{n})$ and 
$(q_{n+1}\alpha - p_{n+1})$ 
have opposite signs.
\end{proof}

\noindent 
\section{The Main Theorem} \label{sec3}
\par \noindent
For $\alpha \in \reals \setminus \rats$ and 
$m \in \nats$, the fractional parts, $\{\alpha\}, \{2\alpha\}, \ldots, 
\{m \alpha\}$, define a partition, $P_{\alpha}(m)$, 
of $[0,1]$:
$$0 = d_{0} < d_{1} < \cdots < d_{j} < 
d_{j+1} < \cdots < d_{m} < d_{m+1} = 1$$

\noindent The maximum and minimum lengths of the 
subintervals of $P_{\alpha}(m)$ are denoted, respectively, by 
\begin{eqnarray*}
	d_{\alpha}(m) & := & \max_{0 \leq i \leq m}
	\left(d_{i+1} - d_{i} \right) \\
	d'_{\alpha}(m) & := & \min_{0 \leq i \leq m}
	\left(d_{i+1} - d_{i} \right) 
\end{eqnarray*}

\par \noindent For the partition $P_{\alpha}(m)$, the 
differences $(d_{i+1} - d_{i})$ can be completely 
characterized [1] in terms of 
$\eta_{n} \equiv |q_{n}\alpha - p_{n}|$. Collecting the
relevant results in reference [1], we have
\newpage
\addtocounter{theorem}{-1}
\begin{theorem}[Three Distance Theorem] \label{thm1}
Let $\alpha \in \reals \setminus \rats$ and $m \in \nats$.  
\begin{itemize}
\item[(a)]  $m$ can be uniquely represented as 
$m = rq_{k} + q_{k-1} + s$, for some $k \geq 0$, $1 \leq r 
\leq a_{k+1}$, and $0 \leq s < q_{k}$ (where $a_{k}$'s are 
the partial quotients of $\alpha$ and $q_{k}$'s are 
given in {\bf Fact \ref{fa2}}).
\item[(b)]  For the partition $P_{\alpha}(m)$, 
there are $[(r-1)q_{k} + q_{k-1} + s + 1]$ 
subintervals of length $\eta_{k}$, 
$[s + 1]$ subintervals of length $\eta_{k-1} - r \eta_{k}$, 
and $[q_{k} - (s+1)]$ subintervals of length 
$\eta_{k-1} - (r-1)\eta_{k}$, where the unique integers 
$k$, $r$ and $s$ are as in part (a). 
\end{itemize}
\end{theorem}

\begin{remark} \label{rem1}
From {\bf Theorem \ref{thm1}}, we observe 
\begin{itemize}
\item[(a)]  $\eta_{k-1} - r\eta_{k} = \eta_{k+1} + 
(a_{k+1}-r)\eta_{k}$, by {\bf Lemma \ref{lem1}}
\item[(b)]  $\eta_{k-1} - (r-1)\eta_{k} = \eta_{k} + 
(\eta_{k-1}-r\eta_{k})$
\item[(c)]  When $q_{k}=s+1$, there are no subintervals 
of length $\eta_{k-1} - (r-1)\eta_{k}$.
\end{itemize}
\end{remark}

\addtocounter{theorem}{-1}
\begin{corollary} \label{cor1}
For $m \in \nats$ and $\alpha \in \reals \setminus \rats$, 
the maximum length, $d_{\alpha}(m)$, and minimum length, 
$d'_{\alpha}(m)$, of the subintervals of partition 
$P_{\alpha}(m)$, are given by: 
\par \noindent {\sf (a)}  When $q_{k} > s+1$, 
\begin{eqnarray*}
	d_{\alpha}(m) & = & \left\{ 
\begin{array}{cl}
	\eta_{k+1} + \eta_{k} & , ~~ r = a_{k+1}  \\
	\eta_{k+1} + (a_{k+1}-r+1)\eta_{k} & , ~~ r < a_{k+1}
\end{array} \right. 
\end{eqnarray*}
\par When $q_{k} = s+1$, 
\begin{eqnarray*}
	d_{\alpha}(m) & = & \left\{ 
\begin{array}{cl}
	\eta_{k} & , ~~ r = a_{k+1}  \\
	\eta_{k+1}+(a_{k+1}-r)\eta_{k} & , ~~ r < a_{k+1}
\end{array} \right. 
\end{eqnarray*}
\par \noindent {\sf (b)}  For all $q_{k} \geq s+1$,
\begin{eqnarray*}
	d'_{\alpha}(m) & = & \left\{ 
\begin{array}{cl}
	\eta_{k+1} & , ~~ r = a_{k+1}  \\
	\eta_{k} & , ~~ r < a_{k+1}
\end{array} \right.
\end{eqnarray*}
\noindent where $k$, $r$, $s$, $a_{k}$, and $\eta_{k}$ 
are as in {\bf Theorem \ref{thm1}}.
\end{corollary}

\begin{proof}
From {\it Remark \ref{rem1}(a)} and {\bf Lemma \ref{lem1}} 
we have, 
$$\eta_{k-1}-r\eta_{k} = 
\left\{ \begin{array}{lll}
    \eta_{k+1} & < & \eta_{k} ~~~, ~~ r = a_{k+1} \\
    \eta_{k+1}+(a_{k+1}-r)\eta_{k} & > & \eta_{k} ~~~, ~~ 
      r < a_{k+1} 
  \end{array} \right.$$
\noindent The corollary follows from 
{\bf Theorem \ref{thm1}}, {\it Remark \ref{rem1}(b)} 
and {\it Remark \ref{rem1}(c)}.
\end{proof}

\begin{theorem}[Main Theorem] \label{thm2} Let $\alpha \in 
\reals \setminus \rats$, $m \in \nats$, and 
$d_{\alpha}(m)$, $d'_{\alpha}(m)$ be, 
respectively, the maximum and minimum lengths of the subintervals 
of the partition $P^{\alpha}(m)$.  The sequence 
$\displaystyle{\left(\frac{d_{\alpha}(m)}
{d'_{\alpha}(m)} \right)_{m=1}^{\infty}}$ is bounded if and 
only if $\alpha$ is an irrational number of constant type.
\end{theorem}

\begin{proof}
$m = rq_{k} + q_{k-1} +s$, where $k$, $r$, and $s$ are the 
unique integers given by {\bf Theorem \ref{thm1}}.
From {\bf Corollary \ref{cor1}} and 
{\bf Lemma \ref{lem1}}, 
$$\displaystyle{\frac{d_{\alpha}(m)}{d'_{\alpha}(m)}} =  
	   \left\{ 
	     \begin{array}{ll}
	     \epsilon + \displaystyle{\frac{\eta_{k+2}}{\eta_{k+1}}} + 
	       a_{k+2} & , ~~ r = a_{k+1} \\
	       & \\
	     \epsilon + \displaystyle{\frac{\eta_{k+1}}{\eta_{k}}} + 
	      (a_{k+1}-r) & , ~~ r < a_{k+1}
	     \end{array} \right. $$
\noindent where $\epsilon = 1$ for $q_{k}>s+1$ and  
$\epsilon = 0$ for $q_{k}=s+1$. \\

\par \noindent {\bf (a)}  If $\alpha$ is of constant type ({\bf 
Definition \ref{def1}}), then the partial quotients, $a_{n}$, 
of $\alpha$, satisfy $a_{n} \leq B(\alpha) < \infty$ for all 
$n \geq 1$.  Since $\displaystyle{\frac{\eta_{j+1}}
{\eta_{j}} < 1}$ for all $j \geq 0$ (by {\bf Lemma \ref{lem1}}), 
$\displaystyle{\frac{d_{\alpha}(m)}{d'_{\alpha}(m)} < 
B(\alpha) + 2}$ for all $m \in \nats$.  Hence, 
$\displaystyle{\left(\frac{d_{\alpha}(m)}{d'_{\alpha}(m)}
\right)_{m=1}^{\infty}}$ is bounded. \\

\par \noindent {\bf (b)}  Suppose 
$\displaystyle{\frac{d_{\alpha}(m)}
{d'_{\alpha}(m)} < B_{0}}$ where $0 < B_{0} < \infty$ 
for all $m \in \nats$.  In particular, 
for $m = q_{k+1}$ [corresponding to $r=a_{k+1}$, $s=0$], 
$\displaystyle{\frac{d_{\alpha}(q_{k+1})}
{d'_{\alpha}(q_{k+1})} 
= \epsilon + \frac{\eta_{k+2}}{\eta_{k+1}} 
+ a_{k+2} < B_{0}}$ for all 
$k \geq 0$.  Hence, $a_{k+2} < B_{0}$ for all $k \geq 0$.  
Setting $B = \max\{B_{0}, a_{1}\}$, 
we have $a_{n} \leq B$ for all $n \geq 1$, and hence 
$\alpha$ is of constant type.
\end{proof}
\vskip 1.5cm
\par \noindent {\bf Acknowledgments}\\
\noindent We would like to thank Professor Paul Zweifel, Virginia 
Tech, for his encouragement and stimulating questions, which led, 
in part, to the present work.  We would also like to thank Robin 
Endelman, Department of Mathematics, Virginia Tech, for many 
helpful suggestions and discussions.\\
\newpage
\par \noindent {\bf References}
\begin{itemize}
\item[[1]]  N.B. Slater, {\sl Gaps and steps for the 
sequence $n\theta$ mod 1}, Proc. Camb. Phil. Soc., 
{\bf 63} 1115-1123 (1967)
\item[[2]]  J. Shallit, {\sl Real Numbers with Bounded 
Partial 
Quotients:  A Survey}, Enseign. Math., {\bf 38} 151-187 
(1992) 
\item[[3]]  G. Karner, {\sl On Quantum Twist Maps and 
Spectral Properties of Floquet Operators}, Ann. Inst. H. 
Poincar\'{e} A, to be published
\item[[4]]  S. Drobot, {\sl Real Numbers}, 
Prentice-Hall, Inc., Englewood Cliffs, N.J., 1964
\end{itemize}

\end{document}